\documentclass[aps,prl,twocolumn,showpacs,superscriptaddress,groupedaddress,footinbib]{revtex4-1}
\usepackage[dvips]{graphicx}
\usepackage{subfigure}
\usepackage{epstopdf}
\usepackage{color}
\usepackage{ucs}
\usepackage{float}
\usepackage[utf8x]{inputenc}

\begin{document}

\title{Simulating Raman Spectra using molecular dynamics, and
  identification of high-pressure phases III and IV in hydrogen} \date{\today}
\author{Ioan B. Magdău}
\author{Graeme~J.~Ackland} 
\affiliation{CSEC, SUPA, School of Physics and Astronomy, The University of Edinburgh,
  Edinburgh EH9 3JZ, UK}

\begin{abstract}
 We present a technique for extracting Raman intensities from \textit{ab
  initio} molecular dynamics (MD) simulations at high temperature.  The
  method is applied to the highly anharmonic case of dense hydrogen up to 500 K for
  pressures ranging from 180 GPa to 300 GPa. On heating or pressurizing we find first-order
  phase transitions at the experimental conditions of the phase III - IV boundary.
   Direct comparison of Raman vibrons with
  experiment provides excellent discrimination between subtly
  different structures, found in MD. We find candidate structures 
  whose Raman spectra are in good agreement with experiment. The new phase obtained in high temperature simulations adopts a dynamic, simple 
  hexagonal structure with three layer types: freely rotating hydrogen molecules, static hexagonal trimers and rotating hexagonal trimers.
  We show that previously calculated structures for phase IV are inconsistent with experiment, and their
  appearance in simulation is due to finite size effects.
\end{abstract}
\maketitle
There have been some notable recent successes of using total
energy calculations based on density functional (DFT) to calculate
expected signals from candidate structures, for comparison with
inconclusive experimental data.  Agreement provides validation of the
DFT structure, and this combined approach can yield more information,
with higher reliability, than either technique alone.

Raman spectroscopy provides one such experimental probe, applicable in
extreme conditions but providing insufficient data to determine
crystal structure or identification of the vibrational
mode \cite{LiLD,MarquesLi}. Reliable calculation of Raman frequencies
and intensities of mechanically stable structures can be obtained
using density functional perturbation theory
(DFPT) \cite{ramanDFPT,dfpt} based on \textit{ab initio} lattice dynamics (LD) \cite{mcw,dfptcastep}.  However
these methods do not include high temperature effects, and fail for
dynamically-stabilized structures with imaginary phonon frequencies.
One solution to this is to extract vibrational frequencies from
molecular dynamics (MD) data \cite{Wang1990,Pinsook,Thomas,Fultz}.

For simple structures this is relatively straightforward; \textit{bcc} titanium
and zirconium being nice examples.  In these materials the soft $T_{1N}$
phonon eigenvector is well defined, and its frequency and width can be
calculated from projection of the MD (or Monte Carlo) trajectories
onto the relevant mode eigenvector, followed by Fourier Transform
(FT) \cite{Pinsook,Souvatzis}.

In lower-symmetry molecular systems there may be many modes which are
formally Raman active, and the coupling between lattice and molecular
modes is typically highly temperature-dependent.  Worst of all are
plastic crystal phases where the molecules can reorient in MD and the
eigenvectors calculated from perturbation theory become totally
irrelevant.

In this Letter we present a method for calculating Raman frequencies
from molecular dynamics, and apply it to the particularly awkward and
topical case of  the
high-frequency vibron modes in high pressure hydrogen.


Although liquid and solid phases I, II of hydrogen have been well studied using MD \cite{ScandaloLiquid,KohanoffPII,KohanoffPIII},
much interest recently has focussed on pressures around
200-300 GPa where several phases are reported.  Generally accepted are a
low temperature phase III \cite{Mao1994} and a high temperature phase
IV \cite{Howie2012,Pick2012a}.  Theoretical predictions of many other phases \cite{Pick2006,Pick2007} have
been made, and Raman data suggests phase IV may itself 
have a subtle structural change at 270 GPa \cite{HowiePRB}.

At these pressures x-ray experiments are exceedingly difficult, while neutron
diffraction is simply impossible; therefore most experimental data are
extracted from Raman and infrared spectroscopy, alongside
conductivity measurements.  None of these techniques produce enough
data to resolve crystal structures, so DFT studies have also been attempted
 \cite{Pick2006,Pick2007,Pick2012a}.  Although these
calculations typically ignore quantum effects on the protons, they
still provide a useful indication of the likely structures.  Very
recent papers \cite{Li,Morales} applying path integral MD to high
pressure hydrogen show no qualitative behavioural change to the phase
diagram: the main effect is that tunnelling allows molecular rotations
to occur at slightly lower temperatures than in classical MD, lowering
the phase lines.  Most importantly for the present work, the
vibrational frequencies of the molecules are largely unchanged by the
path integral dynamics.

\begin{figure*}
\fbox{\includegraphics[width=160mm]{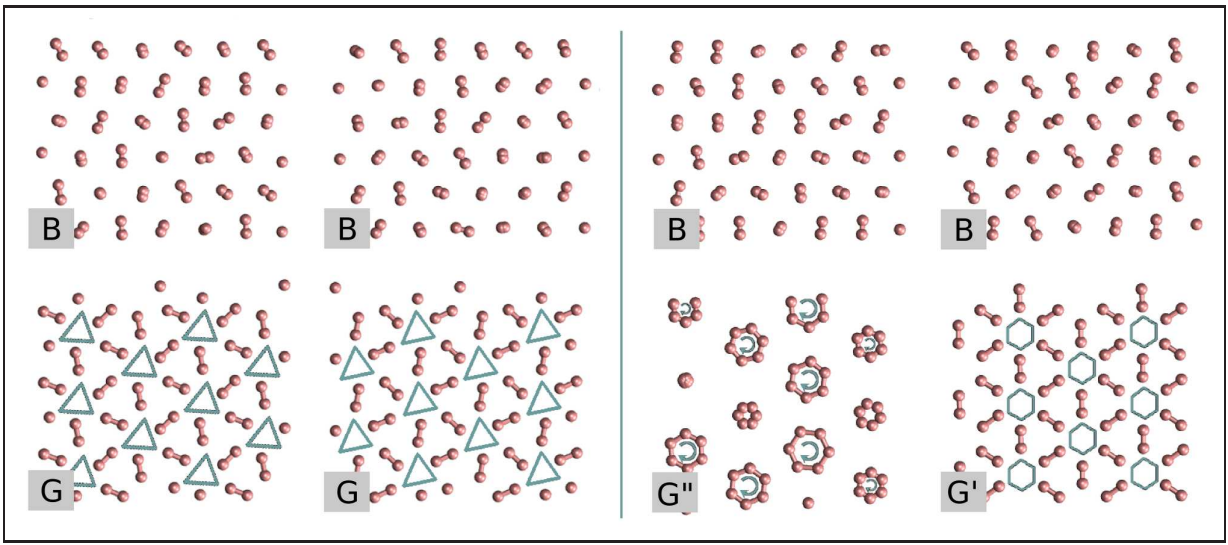}}
\caption{Time-averaged (1ps) atomic positions from simulations
  at pressure/temperature/initial configuration. Left column: phase IVa, stacked
  $BGBG$, 220 GPa /220 K/ Pc; right column: phase IVb, stacked $BG''BG'$, 270 GPa/ 220 K/ Pc.  The time
  averaging is chosen large enough to capture the rotation of the $B$-layer molecules
  and $G''$-layer motifs making these smaller, but not so large that
  the motifs become points at the central position (representing 2 ($B$) and 6
  ($G''$) atoms respectively). Note the strong 6-fold symmetry
  in phase IVb compared with IVa.  Many other similar figures are
  given in \cite{SUP}, showing the trends with P and T.}
\label{phaseIV}
\end{figure*}

Enthalpy is easily calculated in DFT, being the combination of total
energy of binding of electrons to atoms, plus the zero point energy.
Phase III should be the lowest enthalpy phase over a wide range
of pressures.
\textit{Ab initio} structure 
search \cite{Pick2007}  unveiled a number of candidate phases with low total
energy, and evaluation of normal modes, phonon frequencies and zero
point energy gave rise to prediction of a $C2/c$ symmetry phase.
From the phonon calculation it is further possible to calculate
polarization and polarizability, from which Raman and infrared
intensities may be deduced.  These show reasonable agreement between $C2/c$ 
and experiment for phase III \cite{Pick2007,akahama,akahamaX,loubeyre, IRphaseIII}.

Phase IV exists at higher temperatures and is therefore
stabilized by entropy.  
High temperature calculations are more challenging for DFT.  Using the
quasiharmonic approach based on zero-temperature calculations, Pickard
et al \cite{Pick2012a} evaluated free energies at finite temperature to
claim that phase IV should be a layered structure with alternating
graphene-like hexagonal layers interspersed with ordered molecular
layers and $Pc$ symmetry.  We refer to these layers as $G$ and
$B$-type respectively. The critical result here is that the two
different layers give strongly Raman active vibron modes at two very
different frequencies.  Experimental work also finds two vibron modes,
lending support to a two-layer model \cite{Howie2012}.  

\begin{figure}
\begin{center}
\includegraphics[width=70mm]{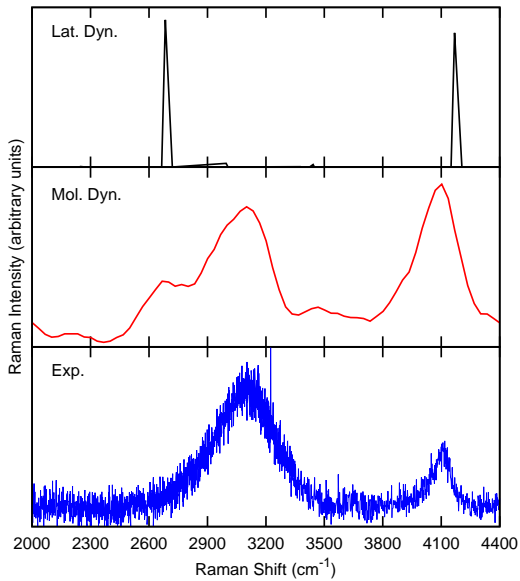}
\end{center}
\caption{ Comparison of Raman spectra at 270 GPa: Top DFPT 
lattice dynamics calculation (0 K); Middle  MD calculation of Raman signal 
from IVb at 220 K; Bottom, experimental data at room temperature from Howie et al \cite{Howie2012}.
  Note that the sensitivity of the detector
  is reduced at high frequency \cite{HowiePRB}, so the peak amplitudes are not
  directly comparable.} 
\label{raman}
\end{figure}

We have extended these calculations of
 enthalpies and phonons using the CASTEP
code \cite{castep} across a wider range of frequencies and with a
variety of pseudopotentials (both ultrasoft and norm-conserving with
various tunings) and exchange-correlation functionals.
We find that these previous results are robust \cite{note1}.  Despite
qualitative agreement, the lower-frequency vibron is observed at much
higher frequency than calculated, and with very large width (Fig.\ref{raman} and \cite{SUP}).
We also used static calculations to investigate whether the $G$ layers have
atomic or molecular character. Mulliken bond analysis
shows very clearly that the $G$-layer hexagonal are rings of three
$H_2$ molecules: the Mulliken charge in the molecular bond is at least
double that between molecules (1.5e vs 0.3e at 180 GPa, closing to 1.3e
to 0.65e at 350 GPa).  This result is consistent with snapshots from MD.
The reduction in molecular fidelity with pressure is accompanied by a steady reduction in the band gap.

We have conducted a range of MD simulations at various conditions of
temperature and pressure, starting in either the $Pc$ or the $C2/c$
phase.  We find that simulations with 48 atoms are plagued with finite size effects
for reasons explained in \cite{SUP}, so all data here come from 288-atom
calculations.
The unit cells of $Pc$ and $C2/c$ are sufficiently dissimilar that
transformation between the two does not occur.  
This makes it clear that MD alone cannot be relied upon to find the experimental structure: comparison with data is essential for validation.
However, we do observe direct
phase transformation in the MD between layered structures with and
without the $B$-type free rotating molecular layer.

In simulations starting in $C2/c$ we find a stable $G$-layered structure at
temperatures up to around 300 K, above which a transformation occurs to a structure 
with alternating $B$ and
$G$ layers similar to the monoclinic C2 \cite{Pick2007}. The transformation mechanism is such that the $GBGB$-structure 
layers become almost orthogonal to the original $C2/c$ \cite{SUP}.  

For simulations starting in $Pc$, we find reversible transitions
between two phases: a structure with $BGBG$ stacking similar to $Pc$,
with threefold layer-symmetry but with the $B$-layer molecules
rotating about their centres, and a high-temperature structure with
hexagonal symmetry, stacked $BG'BG''$ with sixfold layer-symmetry where
the $G'$ layer has hexagonal symmetry and the $G''$ layer exhibits fast
rebonding which enables rotation of the hexagonal motifs of three
hydrogen molecules.  We refer to these phases as IVa and IVb
(Fig.\ref{phaseIV}).  This transition is observed in two ways \cite{SUP}: a single MD run with a ramped temperature rise and subsequent
decrease traverses a path IVa - IVb - IVa with little hysteresis;
alternately, calculations at fixed T and P show the two phases.  We use these
latter calculations as the basis for our Raman calculations, to test
the simulated structures against experiment.  It should be noted that
our simulations have only four layers, so more complex stackings may
exist.

Our DFPT-LD calculations show that all modes in the $Pc$
phase are Raman active.  
To compare with experiment we initially tried projecting the MD trajectories onto LD eigenmodes, multiplying by the calculated Raman intensity and taking the Fourier Transform \cite{SUP}.
 This method
relies on the normal mode vectors being invariant over time and with temperature.  At the
lowest temperature, 60 K, the structure remains in the initial
configuration (metastable $Pc$) and the projection method gives a
Raman signal in precise agreement with LD, as it should for harmonic vibrations \cite{SUP}.
 At higher temperatures molecular rotations exchange atom
positions, and this method fails.

Closer inspection shows that the strongly Raman active vibron modes in
all phases involve in-phase stretches of the molecules in the $G$ and
$B$ layers. Therefore we make the ansatz that, independent of
molecular orientation, the Raman-active vibron modes will involve
in-phase stretches.  Extracting the Raman signal from the MD is now
achieved by identifying molecular bondlengths at each step, which
turns out to be always straightforward, and taking the time FT of the
average projection of the velocity over the bondlengths \cite{SUP}.  We note
that this procedure requires well defined molecules, but does not
require us to identify layers \cite{note2}.  This method produces well-defined peaks, and an example of the fit between simulated
and experimental data at 270 GPa is shown in Fig.\ref{raman}.


\begin{figure}
\begin{center}
\includegraphics[width=85mm]{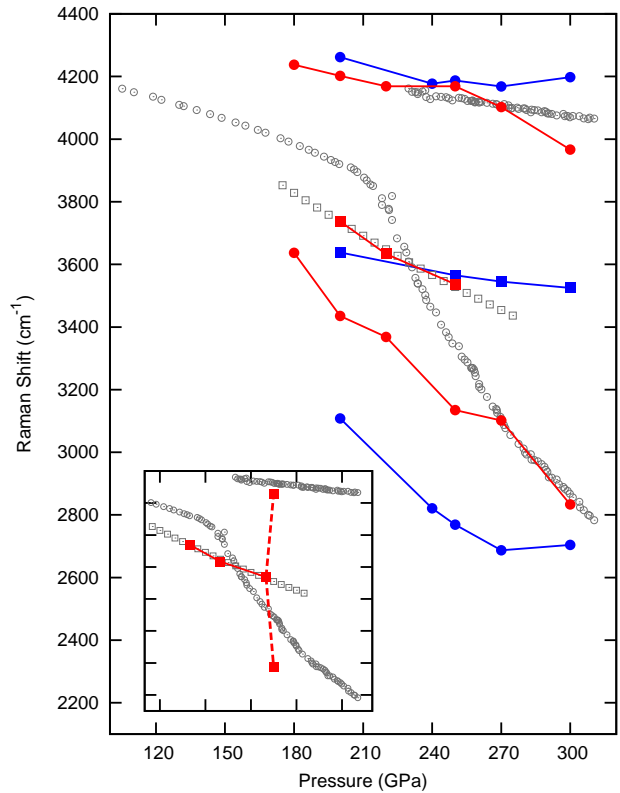}
\end{center}
\caption{Pressure dependence of the vibron peaks.  (a) experiment:
  phase IV, 300 K  \cite{Howie2012}(open grey circles), phase III,
  90 K \cite{akahama}(open grey squares) (b) MD: initialised in $Pc$
  at 220 K (solid red circles) and in $C2/c$ at 220 K (solid red squares)
  (c) Lattice dynamics: in $Pc$ (solid blue circles) and in
  $C2/c$ (solid blue squares). The discontinuity
  in the $Pc$-MD at 270 GPa corresponds to the IVa-IVb transition.
  Inset: the bifurcation of the $C2/c$-MD at 250 GPa corresponds 
 to the two Raman peaks which were calculated after $C2/c$ was heated to 300  K and
 it transformed to a distorted structure similar to phase IV.}
\label{vibronP}
\end{figure}

In Figure \ref{vibronP} we show the pressure dependence of the MD-calculated
Raman vibron frequencies compared with the experimental data.  In
phase III the MD has spectacular agreement with
experiment \cite{akahama}. DFPT gives similar frequencies, but with a
much lower slope (Fig.\ref{vibronP}). The simulation which started in $C2/c$ at 250  GPa/ 220 K was also driven through the phase transition by heating to
300 K, where it formed a frustrated monoclinic, $GBGB$-stacked structure \cite{SUP}
leading to the appearance of a second vibron: this structure has mixed-layer character like phase IV, but the frequencies are clearly not in agreement with experiment (Inset to Fig.\ref{vibronP}).

Figure \ref{vibronP} also shows that the IVb structure with rotating trimers 
does correctly reproduce the experimental frequencies, while IVa and $Pc$
structures do not. The phase IVb lower peak seems to be
comprised of two overlapping peaks(Fig.\ref{raman} and \cite{SUP}). We can project the symmetric
stretch modes layer by layer: for phase IVb with three different
layers, this gives three different frequencies. The near-equal
strength of the $G'$ and $G''$ peaks is due to having one layer of
each in the simulation.  This ratio is determined by finite size, and
in reality the lower peak may be weaker. In fact, this feature is also
probably present as a shoulder in the experiment, although its effect 
could be interpreted as an extended peak width (see Fig. \ref{raman}).  We therefore identify our phase IVb with the
experimentally observed phase.  IVa occurs in a region of PT space
occupied by phase III, so we regard it as metastable.


The analysis of the detailed molecular motions gives a clear, intuitive
picture of the high pressure phase behaviour of hydrogen. At low
temperature we observe a series of $G$ layers to be the stable
structure. Above 60 K molecular motion means that this layer has 3-fold
symmetry, but at 0 K a symmetry-breaking distortion freezes in giving
the $C2/c$ structure.  At a temperature of around 250-300 K
(depending on pressure \cite{SUP}) the transformation to phase IV (our
IVb) occurs.  The explanation for the entropy-driven transition is
evident in the rapid rotational movement of the $B$ layer atoms.  The
molecules rotate rapidly in the $B$ layer such that their
time-averaged positions have hexagonal symmetry.  These rotations
provide the entropy difference between phases III and IV.  At higher
temperature, the fast rebonding in the $G''$ layers enables the rotation of the
hexagonal motifs and adds to the entropy, stabilising the IVb structure.  Once this
rotation begins, the non-rotating $G$ layer adopts hexagonal symmetry.
At lower temperatures, no rotation occurs and the hexagonal motifs are
distorted and symmetry-broken, as in $Pc$.

Our 288-atom results are significantly different from previous MD
work \cite{Gonch2013,Liu} which has been equivocal about the structure
of phase IV, due reorienting of molecules, transformation of one layer
type to another, and ``mixed'' phases of
apparently random $B$ and $G$ layer stacking.  In our own calculations
with 24 or 48 atoms per unit cell we find the same behavior as described in \cite{Gonch2013,Liu}.

A simple 1D model \cite{SUP} of
independent layers shows that the continual layer transformation (random layer stacking)
is to be
expected from finite size effects, rather than phase stability.

Proton ``diffusion'' in the $G$-layers requires two distinct
steps:
\begin{itemize}
\item Rebonding within the rotating hexagonal motifs, a local process which can
  contribute to Raman broadening and may be enhanced by tunnelling.
  This is seen in all simulation sizes at sufficient temperature and illustrated in Figure \ref{phaseIV}.
\item Rearrangement of the
motifs themselves, a process which must occur system-wide, and is seen only in
48 atom simulations and smaller.
\end{itemize}

The rebonding and rotation of these motifs was also described by Liu et al \cite{Liu}, who also
showed large diffusion of hydrogen which implies the rearrangement step.

The 48 atom cell has 2x2 trimer hexagons in a G
layer, and we also observe correlated changes in the identity of pairs in
these motifs, which combined with trimer rotation results in rapid
diffusion of hydrogen through the system.  This effect is not observed
in our 288 atom simulations where
the equivalent mechanism would require correlated changes in four
trimers: we regard the apparent rapid diffusion in 48-atom simulations phase IV 
as a finite size effect \cite{morales2010,Lorenzen}.

Our IVb structure is the best model for the observed phase IV.
Our molecule-based technique shows vibron peaks appear which can be
associated with each layer type.  the B-layer gives the highest
frequency, and analysis of the Raman active modes associated with $G'$
and $G''$ layers in IVb gives two distinct vibrons of slightly different
frequency.  In the overall pattern, these peaks overlap to give a
single peak with a shoulder.  The wide variety of environments in
which the G-layer molecules find themselves leads to a very broad
Raman width. The $B$ layer molecules are well defined and the Raman
peak associated with them is sharper.  In $Pc$, IVa and IVb the
B-layer vibron has similar frequency, however the $G$ layer modes are
quite different.

In sum, we have shown how Raman frequencies can be
extracted from molecular dynamics data allowing direct comparison to
experiment.  We have applied the method to hydrogen at high pressure,
showing that the anharmonicity is so extreme as to invalidate use of
DFPT normal modes, but that in-phase vibrons are the appropriate
coordinates for projection.  Our simulations show several different
phases, some of which are doubtless metastable, but by comparison to
experiment we identify phase III with a structure similar to C2/c, and phase IV
with a high-entropy hexagonal structure of rotating molecules and
trimer motifs.  Our simulations give no support to the notions that
phase IV exhibits either proton transfer, proton tunneling or mixed
molecular and atomic character.

We acknowledge E. Gregoryanz for numerous useful discussions and EPSRC for
a studentship (IBM).

\clearpage

\onecolumngrid

\section{Supplementary Materials}

\subsection{Finite Temperature Phonons by projection onto normal modes}

In order to study a particular vibrational mode in a crystal, we first
define the calculation supercell, and relax the structure at 0 K.  The
atoms are now located at positions given by 3N 
cartesian coordinates $X_j$.  We 
 regard the supercell as a
non-primitive unit cell, in which case ${X}_j$ are the basis
positions. 

We now do a lattice dynamics calculation at 0 K using either 
finite displacements (Ref. \cite{mcw} from the paper) or DFPT (Ref. \cite{dfptcastep} from the paper).  
This gives us a set
of normal mode coordinates ${\bf e}_i$.  With each of these normal
modes we can calculate harmonic phonon frequency ($\omega_i$), Raman
activity and oscillator strength ($R_i$), IR activity and oscillator strength.
$i$ runs from 1 to 3N, the number of normal modes.  all of this is already standard in CASTEP.

From an MD simulation with T timesteps we generate trajectories of the
atoms, ${x}_{i}(t)$, at finite temperature.  We can expand each 
cartesian component of the trajectory 
in terms of the normal modes (ignoring translations).

\[ x_{j}(t) =  X_j+ \sum_{i=4}^{3N}\alpha_{i}(t) {e}_{ij} \]

So far all this is exact, we just made a linear transformation of the
coordinate system.  $\alpha_{i}(t)$ is fully determined by the MD.
Similarly for velocities:

\[ {  \dot{x}}_{j}(t) =  \sum_{i=4}^{3N}\dot{\alpha}_{i}(t) {e}_{ij} \]

Now we assume that we are in the harmonic regime.  
\begin{equation}
\alpha_{i}(t) = Re \left [ a_i\exp^{i(\omega_it+\phi_i)} \right]
\end{equation}
\begin{equation}
\dot{\alpha}_{i}(t) = Im \left [ a_i\omega_i\exp^{i(\omega_it+\phi_i)} \right]
\end{equation}

This assumption means that $a_i$, $\omega_i$ and $\phi_i$ are
independent of time.

It is now straightforward to use the MD data to obtain $\omega_i$,
from the FT.  The FT of $\alpha$,
is problematic since at high temperature $<\alpha>\ne 0$,
but the same
information is present in  $\dot{\alpha}$ which is more convenient since  $<\dot{\alpha}>= 0$.  In the harmonic limit
 $FT[\dot\alpha_i(t)]$ is simply a delta function at $\omega=\omega_i$.  

Note we have NOT used the frequencies from the lattice dynamics, we
have calculated them from the MD. In the harmonic approximation, the
same modes will be Raman/IR Active in the MD as in the lattice
dynamics.  As usual, we can calculate
the occupied phonon density of states from the velocity autocorrelation function:

\[ FT \left [\sum_j \dot{x}_j(t)\dot{x}_j(0) \right ] = FT \left [\sum_i\sum_j 
\dot{\alpha}_i(t)\dot{\alpha}_i(0)e_{ij}^2 \right ] \]

By analogy, the total Raman signal becomes:

\[ FT \left [\sum_{ij} R_i \dot{\alpha}_i(t)\dot{\alpha}_i(0) e_{ij} \right ]\]

and we can obtain the mode frequency for each mode $i$ from the peak  in: $FT \left [ \dot{\alpha}_i(t)\dot{\alpha}_i(0)  \right ]$.

In the harmonic limit, the Raman signal is simply the sum of individual modes. 

All of this has been applied in classical MD by numerous authors, e.g. Ref. \cite{Pinsook} from the paper.  We
now, consider applying exactly the same process to an anharmonic MD.
Modes with strong Raman/IR signals will still have strong Raman
signals, since the polarisability ultimately depends on the motion 
of the atoms.

There are some issues about the magnitude of the oscillations.  In the
harmonic case it will never equilibrate.  To get close to equilibrium
it seems sensible to set initial displacements and velocities based on
temperature from the normal modes (with random phase $\phi$).
\[\alpha_i(t=0)= \sqrt{kT/m\omega_i^2} \sin(\phi)\]
\[\dot{\alpha}_i(t=0)= \sqrt{kT/m} \cos(\phi)\]
This is done, e.g. in SCAILD (Ref. \cite{Souvatzis} from the paper), but not automatically in
CASTEP.  This could be important in evaluating Raman intensities and
line widths, since the anharmonic effects will depend on the phonon
amplitude.  However, for high pressure studies the experimental Raman
intensities depend strongly on the apparatus and are not used
quantitatively.

For H$_2$ vibrons the Raman activity comes from the symmetric
molecular stretch.  It is therefore necessary to associate $\alpha$
not with fixed normal modes, but with the molecule stretches:

\[ \alpha(t) = \sum_{j}{\bf r_j(t) - r_{j_m}(t)} \] 

where ${\bf r}_{j_m}(t) $ is the vector position of the molecular partner atom to $j$, at time t.
 
This requires us to identify molecules at each time step t (i.e. molecule
labelling might change during the simulation, which could cause
discontinuities in the velocity functions). Since the stretching modes can
change at each time step, $\dot{\alpha}(t)$ is not the simple time derivatie of
$\alpha(t)$, but can be calculated by projecting the velocity $v_j(t)$ of each atom
onto the stretching mode. Finally, the spectrum for the vibron modes is: 

\[ FT \left [\ \sum_{j}{\bf v_j(t)\cdot[r_j(t) - r_{j_m}(t)]} \right ]\]

\begin{figure}[H]
\begin{center}
\includegraphics[width=120mm]{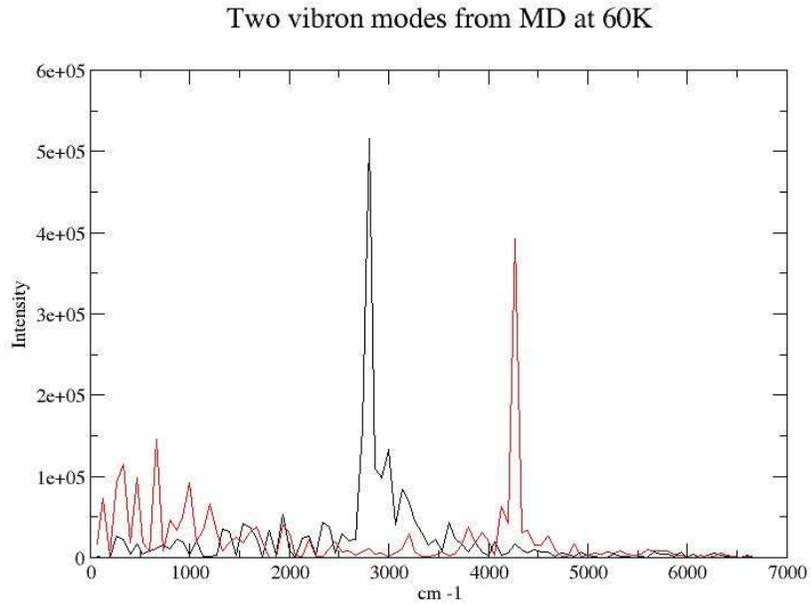}
\end{center}
\caption{Fourier transform of projected velocity autocorrelation
  function of Raman active normal mode.  The two Raman-active vibron
  modes were chosen, and can be seen to be in excellent agreement with
  lattice dynamics values 2845 $cm^{-1}$ and 4187 $cm^{-1}$. From MD calculation on $Pc$ at 250 GPa, 60 K.
  Data in Fig.2 were obtained from peaks in graphs such as this.
\label{Pc}}
\end{figure}

Here we investigate vibrons, but the method is completely general provided
that the Raman-active molecular mode can be identified.
To validate our code, we performed an MD simulation at 60 K starting in
the $Pc$ structure.  Figure \ref{Pc} shows very good agreement in
this simple regime where only harmonic effects are present.  The MD
projection onto normal modes or symmetric stretches give
indistinguishable results.

\subsection{Finite size effects and a simplified layer model for dense H$_2$}

We have seen that the primary feature of the structural heirarchy is
the layer, the secondary feature is the interaction between layers,
and that interactions beyond this are weak.  From the MD we note that
adjacent B layers are not observed, presumably high in energy, while
numerous relative translations of the G layers are observed, depending
on kinetics: this implies little energy preference.

It is possible to understand the transition using a simple 1D model.
We assign differences in energies ($U_{GB}=U_G-U_B<0$ and entropies
$S_{GB}=S_G-S_B<0$ to each layer, and a layer interaction $J_{ij}$
where $i$ and $j$ represent G or B layers.  
In this model the free energies of various N atom supercells with L layers are given in table \ref{layerTable}.

\begin{table}[H]
\centering
\begin{tabular}{|l|l|c|}
\hline
Phase & Stacking & Free energy \\
\hline
III & all-G & $N(U_G-TS_G+J_{GG})$ \\ 
I & all-B & $N(U_B-TS_B+J_{BB})$ \\ 
IV & GBGB &  $N(U_B-TS_B+U_G-TS_G+J_{GB}$)/2\\
& random &  $N(2U_B-2TS_B+2U_G-2TS_G+J_{BB}+J_{GG}+2J_{GB})/4-LT\ln{2}$\\
\hline
\end{tabular}
\caption{Energies of four possible ``phases'' which could be realised in an L=4 layer supercell, where in the general case
L is the number of layers.}
\label{layerTable}
\end{table}

\begin{figure}[H]
\begin{center}
\includegraphics[width=160mm]{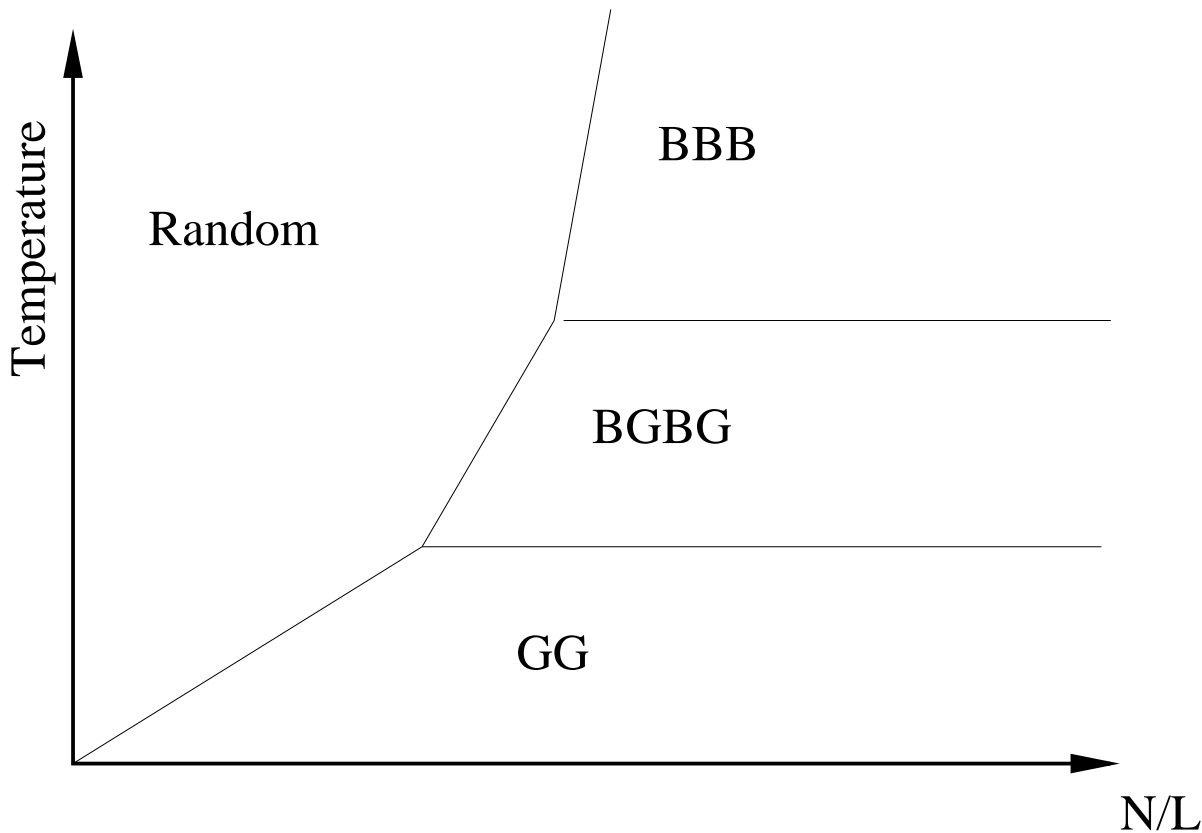}
\end{center}
\caption{Schematic drawing of the effect of finite size and
  temperature for the simple layer model.  $Random$ represents the phase
  observed in simulations with 24 or 48 atoms, $BB$, $BGBG$, and $GGG$ are
  similar to phases III, IV and I respectively.
  For small enough systems
  (low $N/L$) randomly oriented layers will always be stable, however
  in the thermodynamic limit $N/L\rightarrow\infty$ the phase
  sequence increasingly favours free-rotor $B$-layers, the mixed layer
  appearing whenever BB interactions are strongly disfavoured.  The
  actual values of T and N/L at the phase boundary depend on the
  parameters, which in turn depend on pressure and temperature.}
\label{layermodel}
\end{figure}

Assuming $J_{GG}<J_{GB}<J_{BB}$ 
this model gives a phase diagram as shown in figure \ref{layermodel}, including a
III-IV transition and a phase of
all freely rotating molecules (rather similar to phase I).  
As discussed above in the context of 48 atom
simulations, random stacking is favoured by the $LT\ln2$ term, which
is significant only for small system sizes where the number of layers
is comparable to the number of atoms.

Each layer contains only 12 atoms.  For a free energy difference of
$F_{BG}$ per atom between the distinct layer types, Boltzmann
statistics shows that the probability of finding the $B$ layer is
$1/(1+\exp(-12F_{BG}/KT))$.  Static calculation implies a
$F_{BG}\approx 2$meV/atom, so at 300 K the unfavoured layer is present
28\% of the time.

\subsection{Calculation details}

Data was collected from DFT calculations using the CASTEP package.
The MD calculations involved 288 atoms initiated in prerelaxed monoclinic
supercells of either $Pc$ ($\beta\approx 91^{\circ}$) or
$C2/c$ ($\beta\approx 144^{\circ})$ structures, and used a
constant-stress Parrinello-Rahman barostat. Observed phase transitions
involved small changes of cell shape, but nothing close to the
$53^{\circ}$ change required to go from $Pc$ to $C2/c$.

We used the PBE exchange correlation functional which has become the
standard for work in hydrogen \cite{pbe}.  Two different
psuedopotentials were developed, an ultrasoft (300eV cut off)
generated ``on the fly'' \cite{Note3}
for the molecular dynamics and a harder
norm-conserving pseudopotential (1200eV cut off) for which Raman
calculations are more easily carried out.  The structural results
obtained were similar for the two methods.  For the DFPT lattice
dynamics a single unit cell k-point set of 9x5x5 was used, giving 69
independent k-points.

\begin{table}[H]
\centering
\begin{tabular}{|l|ccc|cc|}
\hline
Norm conserving Pc&&&&&\\
P  &a & b&  c&  vibrons & \\
200 & 2.986 & 5.207 & 5.324 & 3108 4262 &\\
220 & 2.949 & 5.129 & 5.259 & 2929  4208 &\\
240 & 2.915 & 5.071 & 5.202 & 2821 4177 &\\
250 & 2.896 & 5.054 & 5.163 & 2845 4187 &\\
270 & 2.868  &4.989 & 5.121 & 2684 4168 &\\
300  &2.826  &4.917  &5.048  & 2597 4157& \\
350&  2.767& 4.815& 4.940
 & 2549   4136 & \\
\hline
USP Pc&&&&&\\
P  &a & b&  c& &\\
180 & 3.031 & 5.272 & 5.405 &&\\
200 & 2.986 & 5.194 & 5.329 && \\
250 & 2.894 & 5.051 & 5.165 && \\
300 & 2.823 & 4.929 & 5.033 &&  \\
\hline
C2c&&&&&\\
P  &a & b&  c& $\beta$ & vibron \\
200 &  5.221663 &  2.976715  &  4.379227 &  142.433942 & 3638.156521 \\ 
230 &  5.126252 & 2.920341& 4.297475  &142.467436 & 3589.690041  \\ 
250 &  5.069649 & 2.887193 & 4.249310& 142.485532 &  3564.867343  \\
270 &   5.017617 & 2.856906  &  4.205190  &  142.501804 & 3545.228906\\ 
300 &  4.9468405 & 2.8159880 & 4.145307 &  142.524403 &  3524.644474 
\\ 
\hline
\end{tabular}
\caption{Structural details from static relaxations of $Pc$ and $C2/c$ structures.  These structures were used for DFPT-LD calculation, and to initialise the MD calculations}
\end{table}

\begin{table}[H]
\centering
\begin{tabular}{|c|c|c|c|c|c|c|c|c|}
\hline
No. & Initial Structure & Atoms & Ensemble & Number of Iterations & Time Step & Pressure & Temperature \\
\hline
 1 & Pc & 48 & NVE & 9000 & 0.5 fs & 250 GPa & 60 K \\
\hline
 2 & Pc & 48 & NVE & 6000 & 0.5 fs & 250 GPa & 145 K \\
\hline
 3 & Pc & 48 & NVE & 6000 & 0.5 fs & 250 GPa & 215 K \\
\hline
 4 & Pc & 48 & NVE & 6000 & 0.5 fs & 250 GPa & 285 K \\
\hline
 5 & Pc & 48 & NVE & 6000 & 0.5 fs & 250 GPa & 360 K \\
\hline
 6 & Pc & 48 & NVE & 6000 & 0.5 fs & 250 GPa & 430 K \\
\hline
 7 & Pc & 48 & NVE & 6000 & 0.5 fs & 250 GPa & 500 K \\
\hline
 8 & Pc & 288 & NVE & 3000 & 0.5 fs & 250 GPa & 145 K \\
\hline
 9 & Pc & 288 & NVE & 3000 & 0.5 fs & 250 GPa & 215 K \\
\hline
10 & Pc & 288 & NVE & 3000 & 0.5 fs & 250 GPa & 285 K \\
\hline
11 & Pc & 288 & NVE & 3000 & 0.5 fs & 250 GPa & 360 K \\
\hline
12 & Pc & 288 & NVE & 3000 & 0.5 fs & 250 GPa & 430 K \\
\hline
13 & Pc & 288 & NVE & 3000 & 0.5 fs & 250 GPa & 500 K \\
\hline
14 & Pc & 288 & NPT + NVE & 500 + 1500 & 0.5 fs & 180 GPa & 220 K \\
\hline
15 & Pc & 288 & NPT + NVE & 500 + 1500 & 0.5 fs & 200 GPa & 220 K \\
\hline
16 & Pc & 288 & NPT + NVE & 500 + 1500 & 0.5 fs & 220 GPa & 220 K \\
\hline
17 & Pc & 288 & NPT + NVE & 500 + 1500 & 0.5 fs & 250 GPa & 220 K \\
\hline
18 & Pc & 288 & NPT + NVE & 500 + 1500 & 0.5 fs & 270 GPa & 220 K \\
\hline
19 & Pc & 288 & NPT + NVE & 500 + 1500 & 0.5 fs & 300 GPa & 220 K \\
\hline
20 & C2/c & 288 & NPT + NVE & 500 + 1500 & 0.5 fs & 200 GPa & 220 K \\
\hline
21 & C2/c & 288 & NPT + NVE & 500 + 1500 & 0.5 fs & 220 GPa & 220 K \\
\hline
22 & C2/c & 288 & NPT + NVE & 500 + 1500 & 0.5 fs & 250 GPa & 220 K \\
\hline
\shortstack{23\\ \hphantom{G} \\ \hphantom{.} } & \shortstack{GGGG \\ stacking} & \shortstack{288\\ \hphantom{G} \\ \hphantom{.}} & \shortstack{NPT\\ \hphantom{G} \\ \hphantom{.}} & \shortstack{ 7 x 200 + \\ 6 x 200 } & \shortstack{0.5 fs\\ \hphantom{G} \\ \hphantom{.}} & \shortstack{250 GPa\\ \hphantom{G} \\ \hphantom{.}} & \shortstack{ \hphantom{G} \\ 100 K $\rightarrow$ 400 K ($+$50 K) \\ 350 K $\rightarrow$ 100 K ($-$50 K)} \\
\hline
24 & C2/c & 288 & NPT & 400 + 600 + 200 & 0.5 fs & 250 GPa & 200 K $\rightarrow$ 300 K $\rightarrow$ 400 K\\
\hline
25 & C2/c & 288 & NPT + NVE & 800 + 2000 & 0.5 fs & 250 GPa & 300 K\\
\hline
\end{tabular}
\caption{Summary of MD calculations.}
\label{MDTable}
\end{table}

\begin{figure}[H]
\begin{center}
\includegraphics[width=150mm]{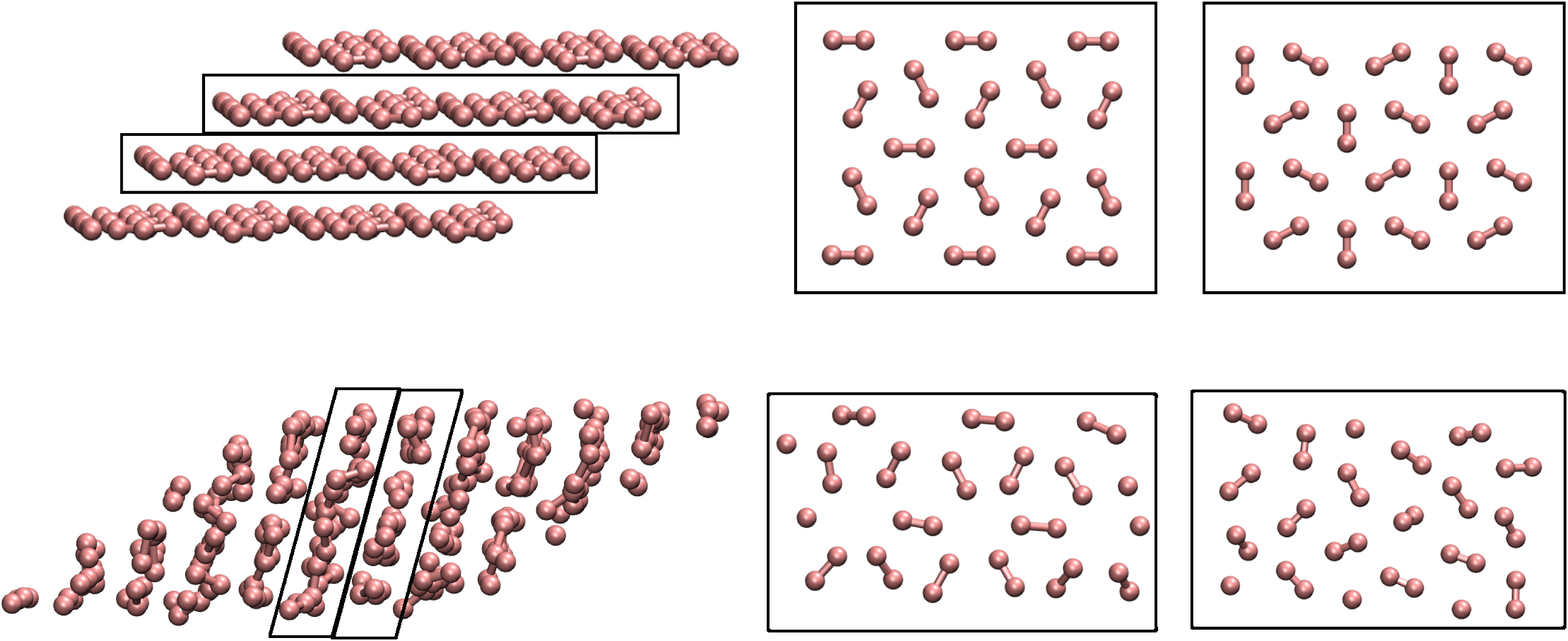}
\end{center}
\caption{Phase III transformation observed in NPT MD initialized in
  $C2/c$ structure and heated at 250 GPa (see simulation 24 from Table \ref{MDTable} and Fig. \ref{heatcool}).  Left side shows MD supercell, while
  right side shows the corresponing layers. Top structure is Pickard's C2/c (Ref. \cite{Pick2007} from the paper)
  relaxed at 250 GPa, while bottom structure is a snapshot from MD, after the transformation
  has occured: the $G$-layer stacking of the $C2/c$ at the
  top is clear.  The alternating $BGBG$ stacking in the high temperature
  phase (lower) is similar to phase IV, but there is some frustration,
  which leads to a lower Raman-mode frequency (around 2980 cm$^{-1}$).
  The frustration illustrates the heirarchical natural of the bonding:
  a primary tendency to form layers, secondary to order as $B$ or $G$
  within the layers, and a tertiary effect of interlayer interactions.}
\label{PTo}
\end{figure}

\begin{figure}[H]     
\begin{center}
\subfigure[. 145 K]{\fbox{\includegraphics[width=75mm]{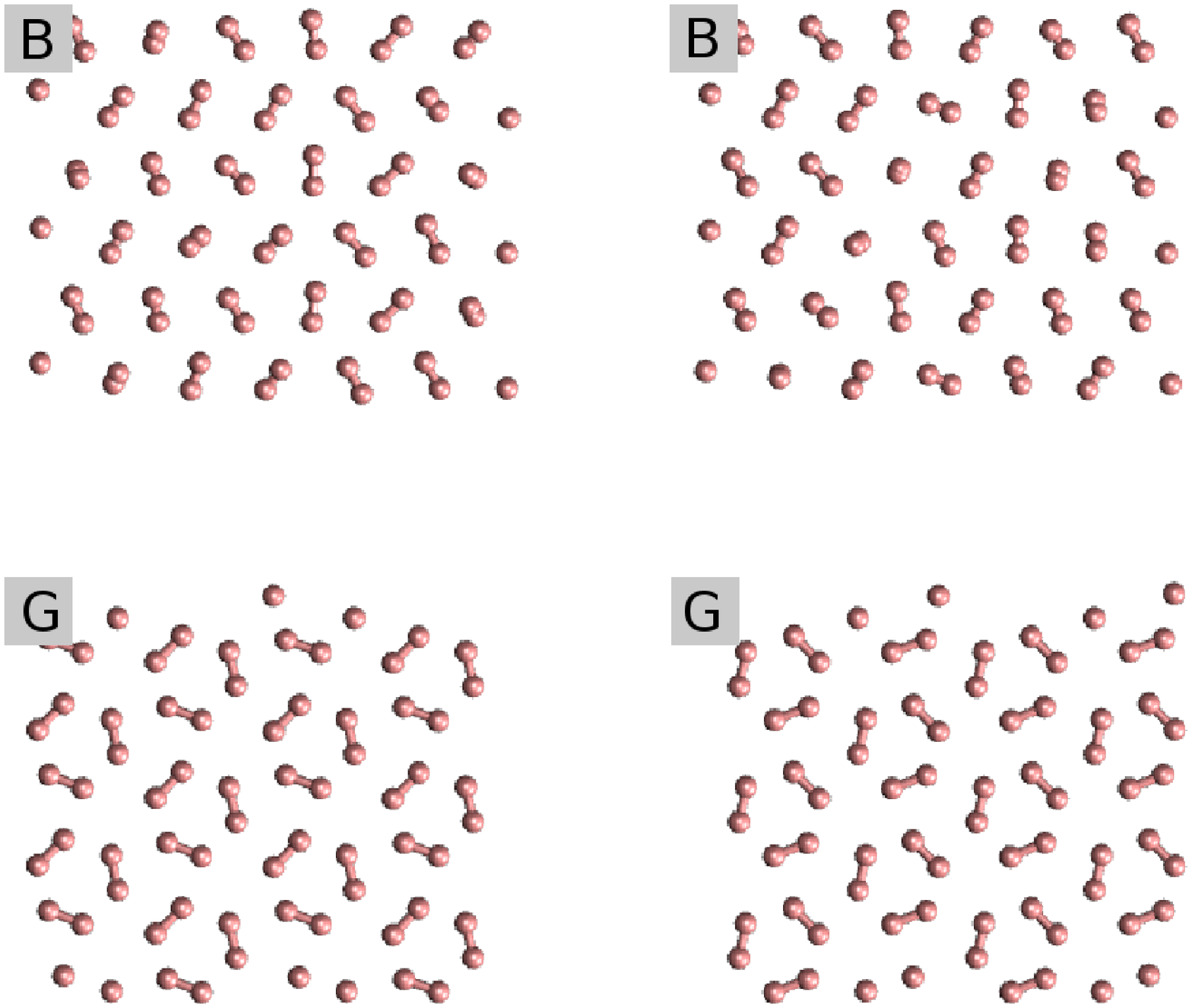}}}
\subfigure[. 215 K]{\fbox{\includegraphics[width=75mm]{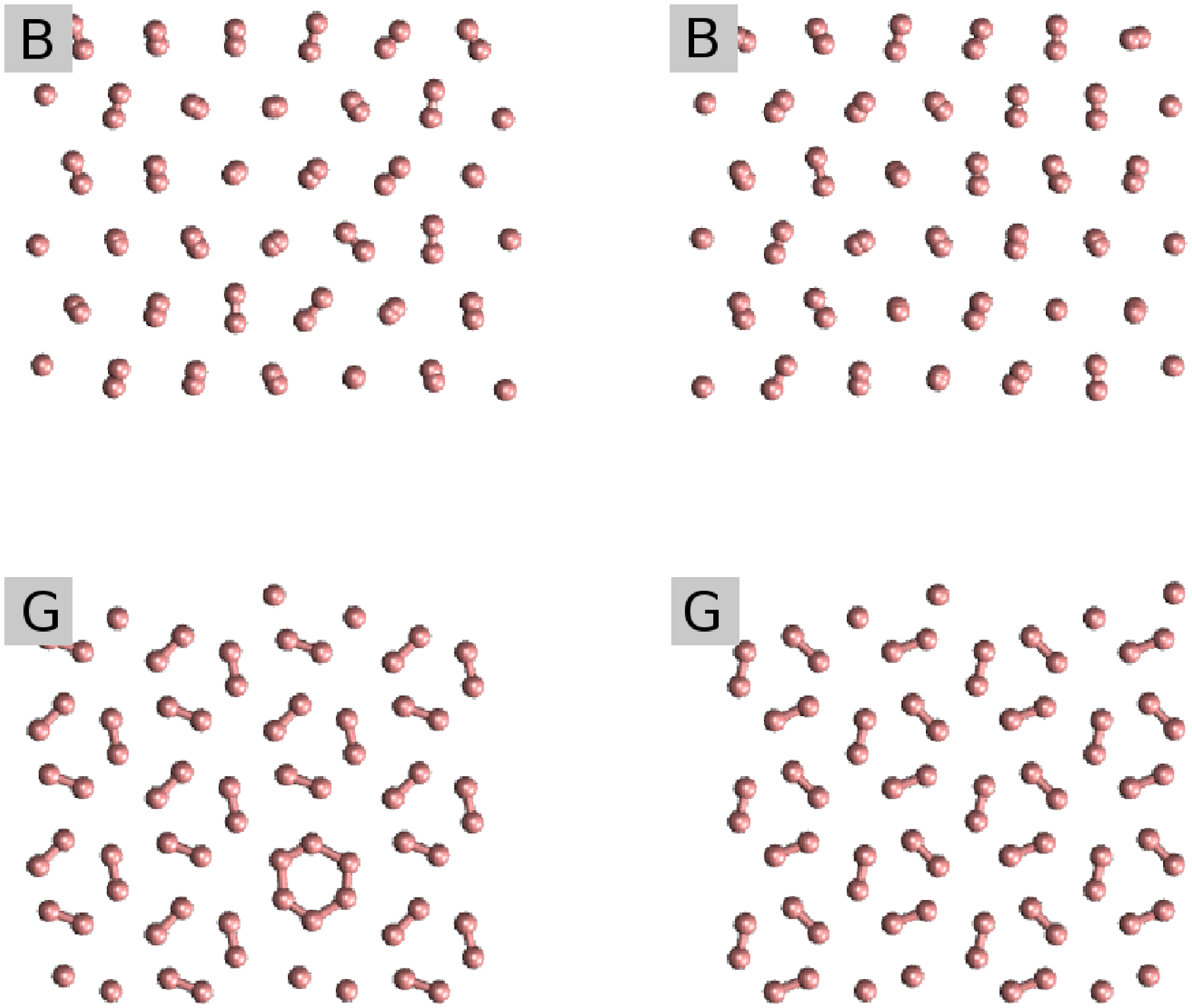}}}
\subfigure[. 285 K]{\fbox{\includegraphics[width=75mm]{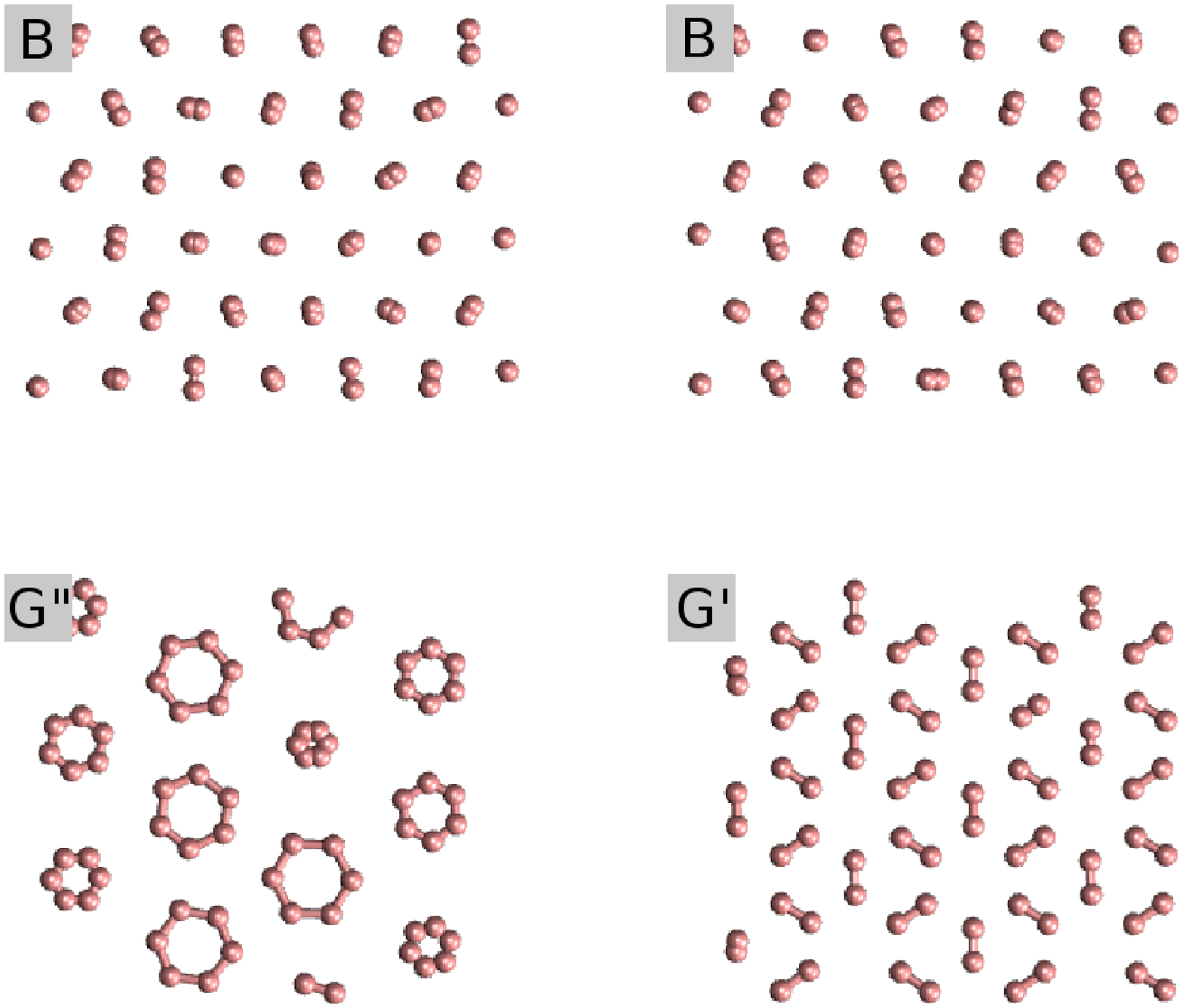}}}
\subfigure[. 360 K]{\fbox{\includegraphics[width=75mm]{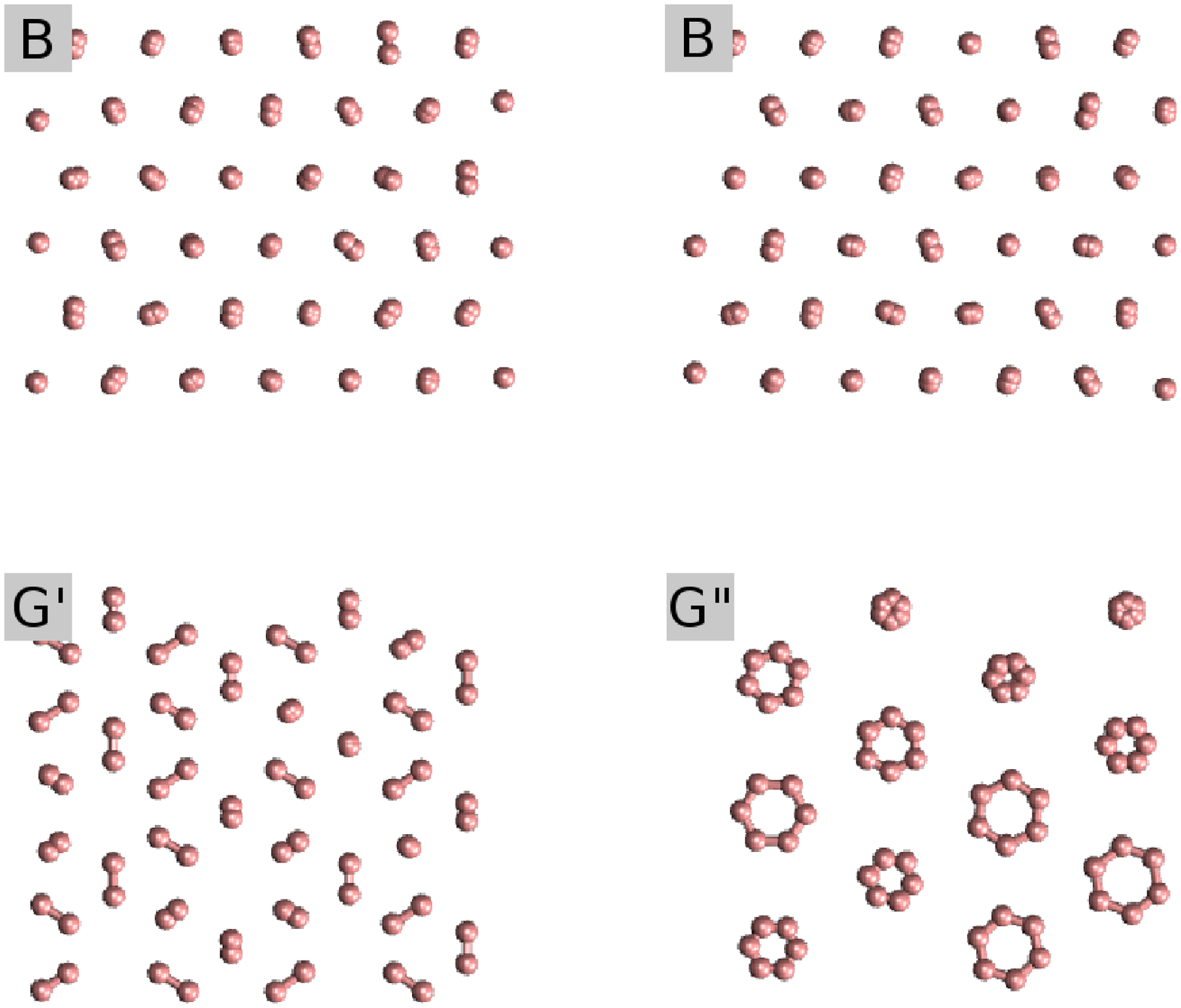}}}
\subfigure[. 430 K]{\fbox{\includegraphics[width=75mm]{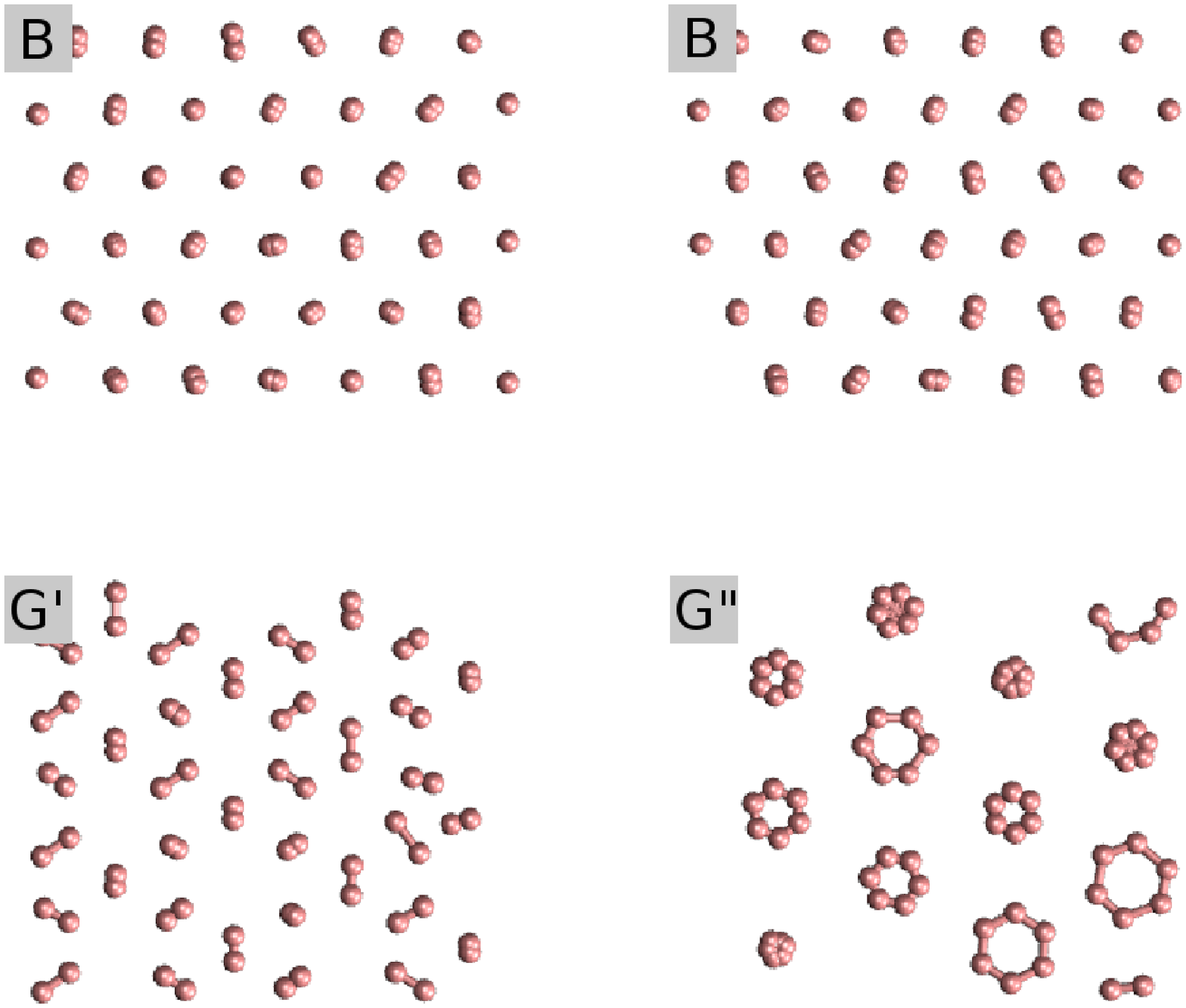}}}
\subfigure[. 500 K]{\fbox{\includegraphics[width=75mm]{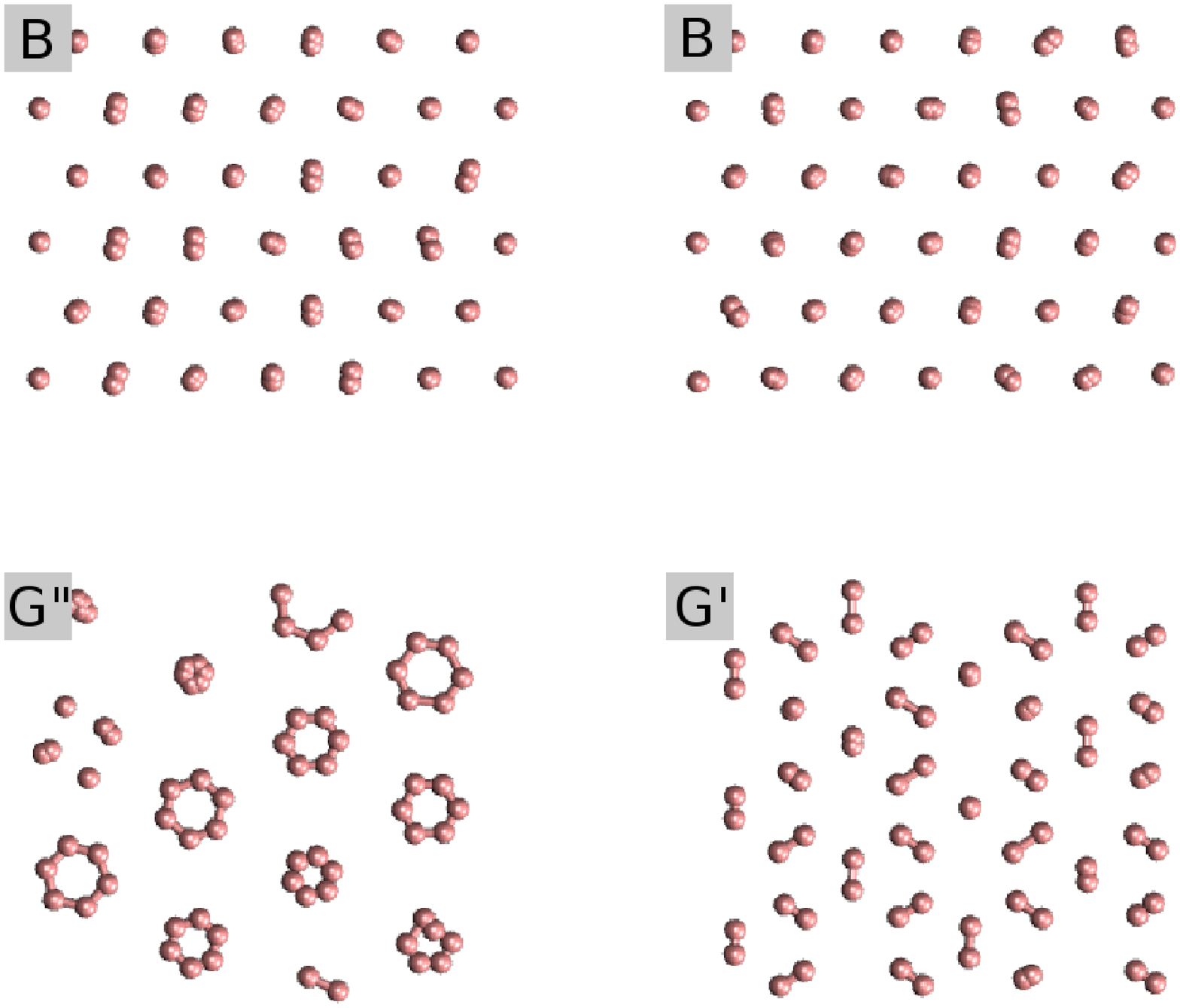}}}
\end{center}
\caption{As per figure 1 in the main paper, average positions of atoms over 1.5ps
at 250 GPa and temperatures (from top left) 145 K, 215 K, 285 K, 360 K, 430 K, 500 K. Stacking is $BGBG$ and $BG'BG''$, repsectively.
\label{phaseIVX_T}}
\end{figure}

\begin{figure}[H]     
\begin{center}
\subfigure[. 180 GPa]{\fbox{\includegraphics[width=75mm]{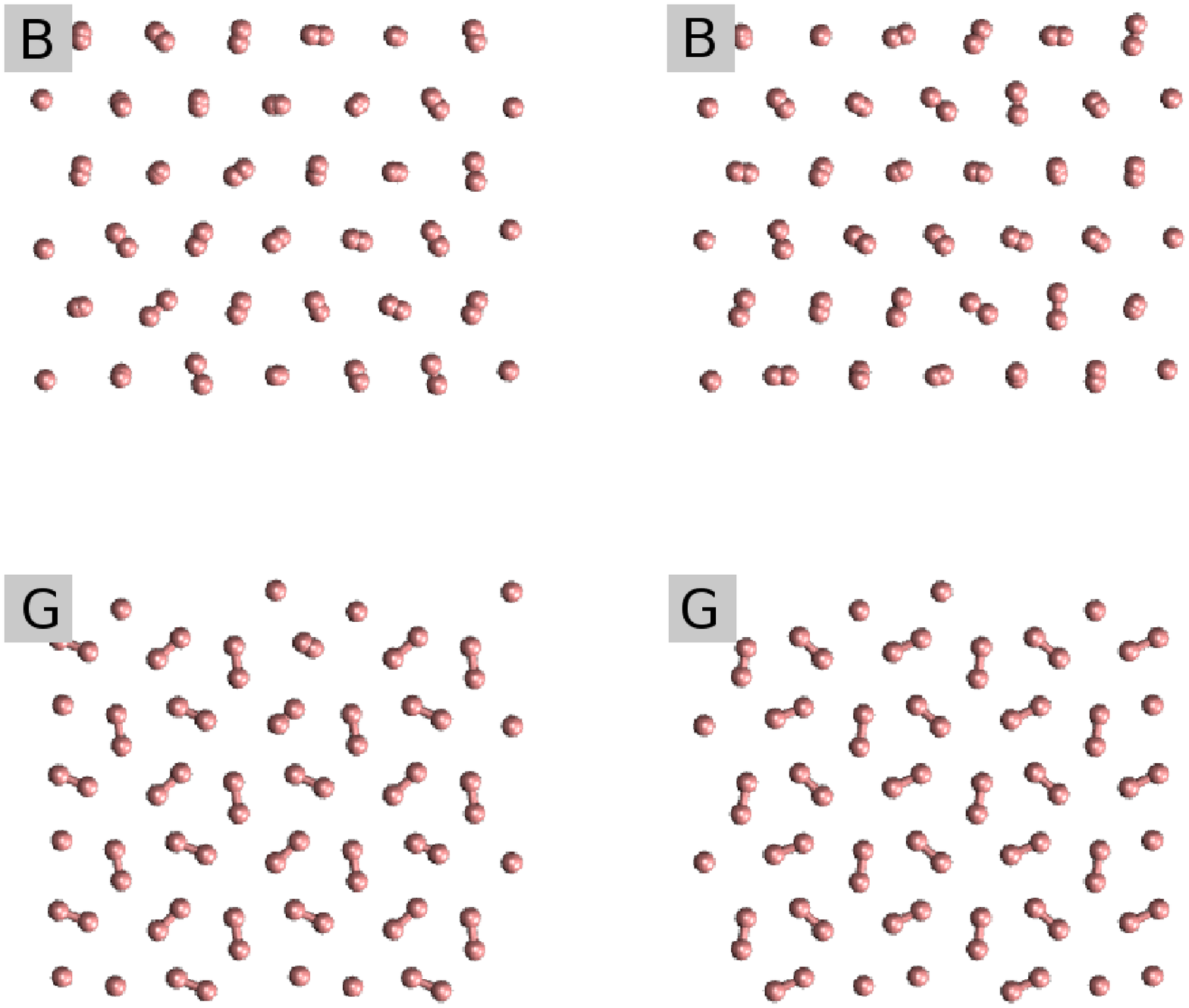}}}
\subfigure[. 200 GPa]{\fbox{\includegraphics[width=75mm]{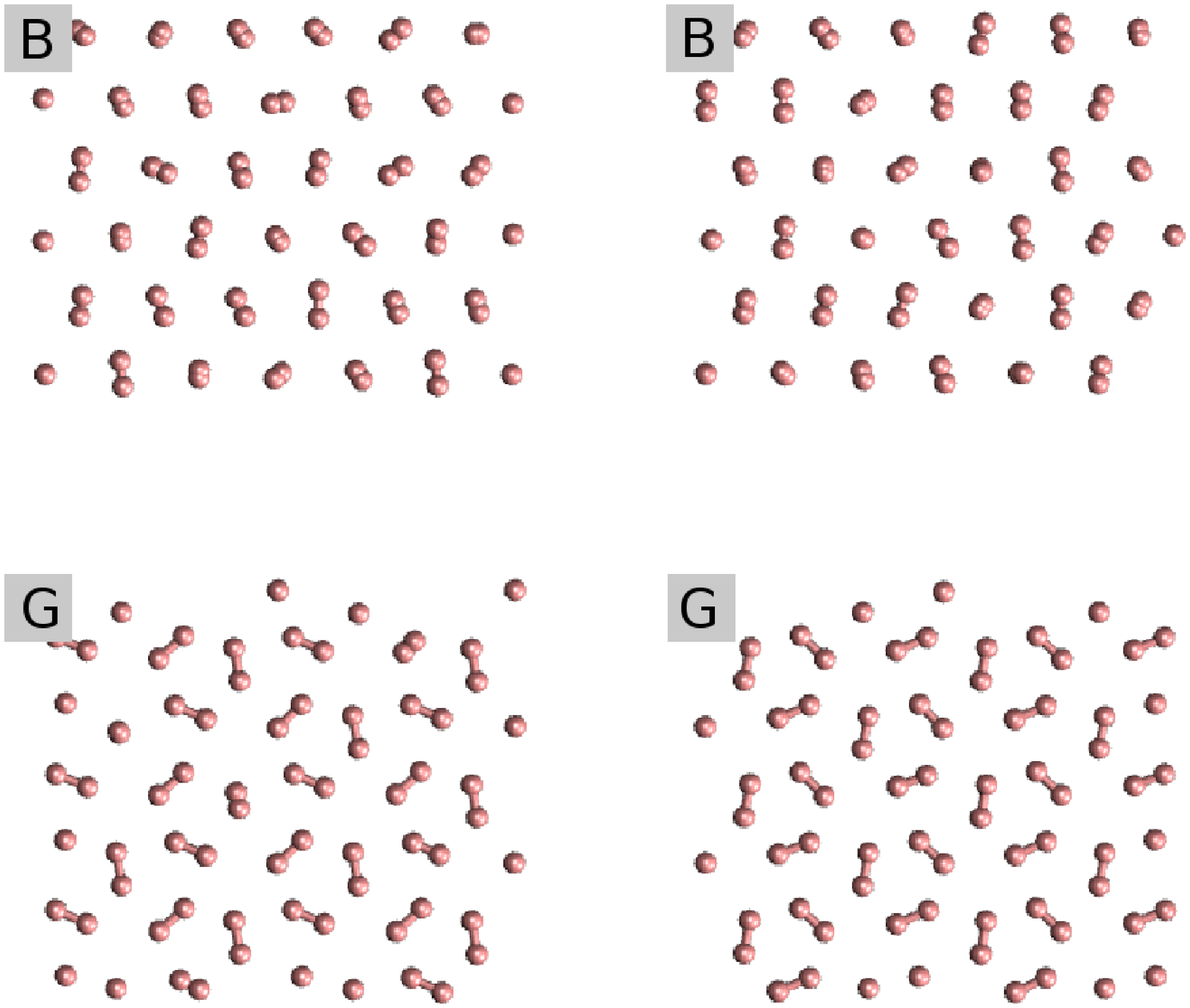}}}
\subfigure[. 220 GPa]{\fbox{\includegraphics[width=75mm]{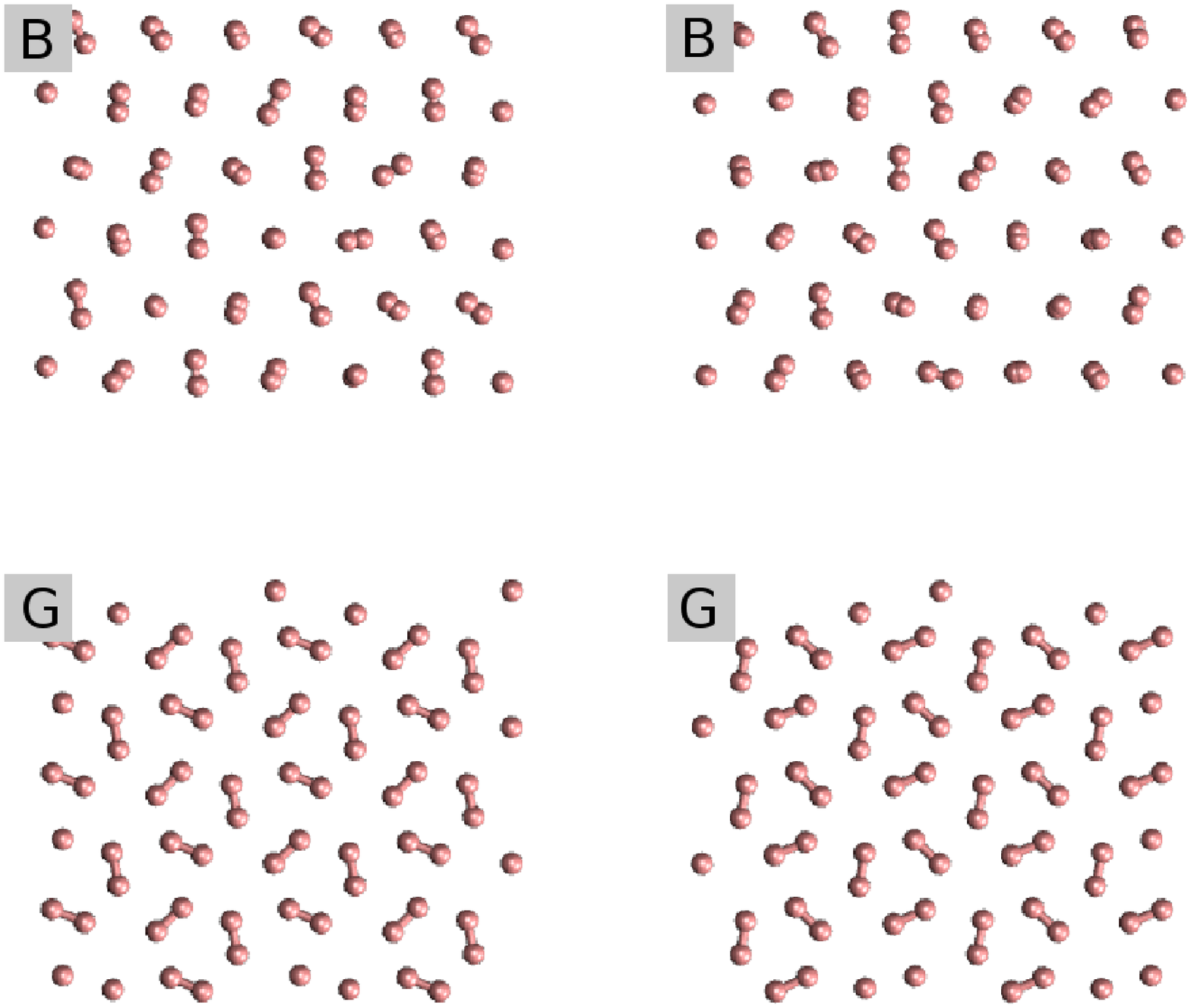}}}
\subfigure[. 250 GPa]{\fbox{\includegraphics[width=75mm]{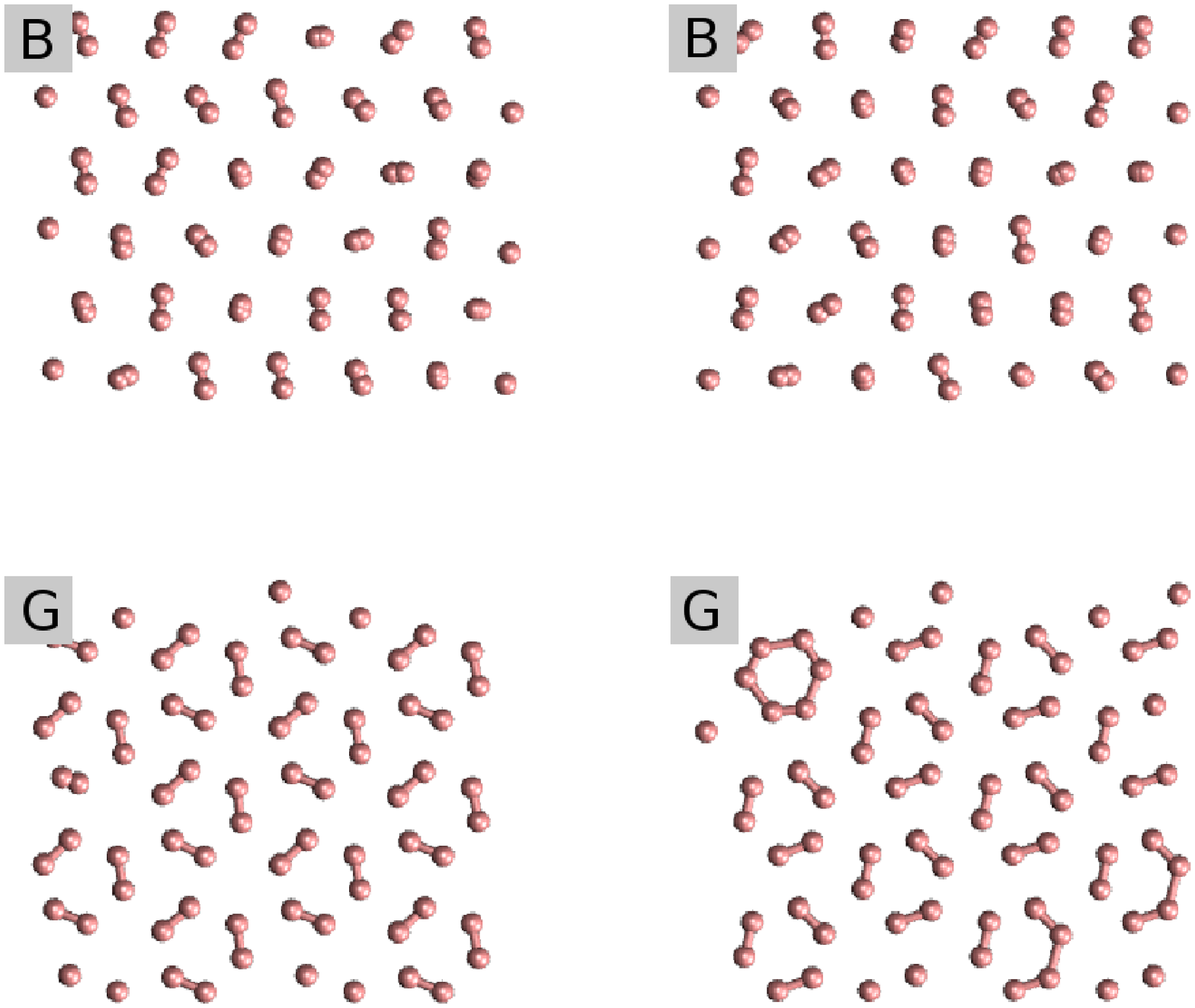}}}
\subfigure[. 270 GPa]{\fbox{\includegraphics[width=75mm]{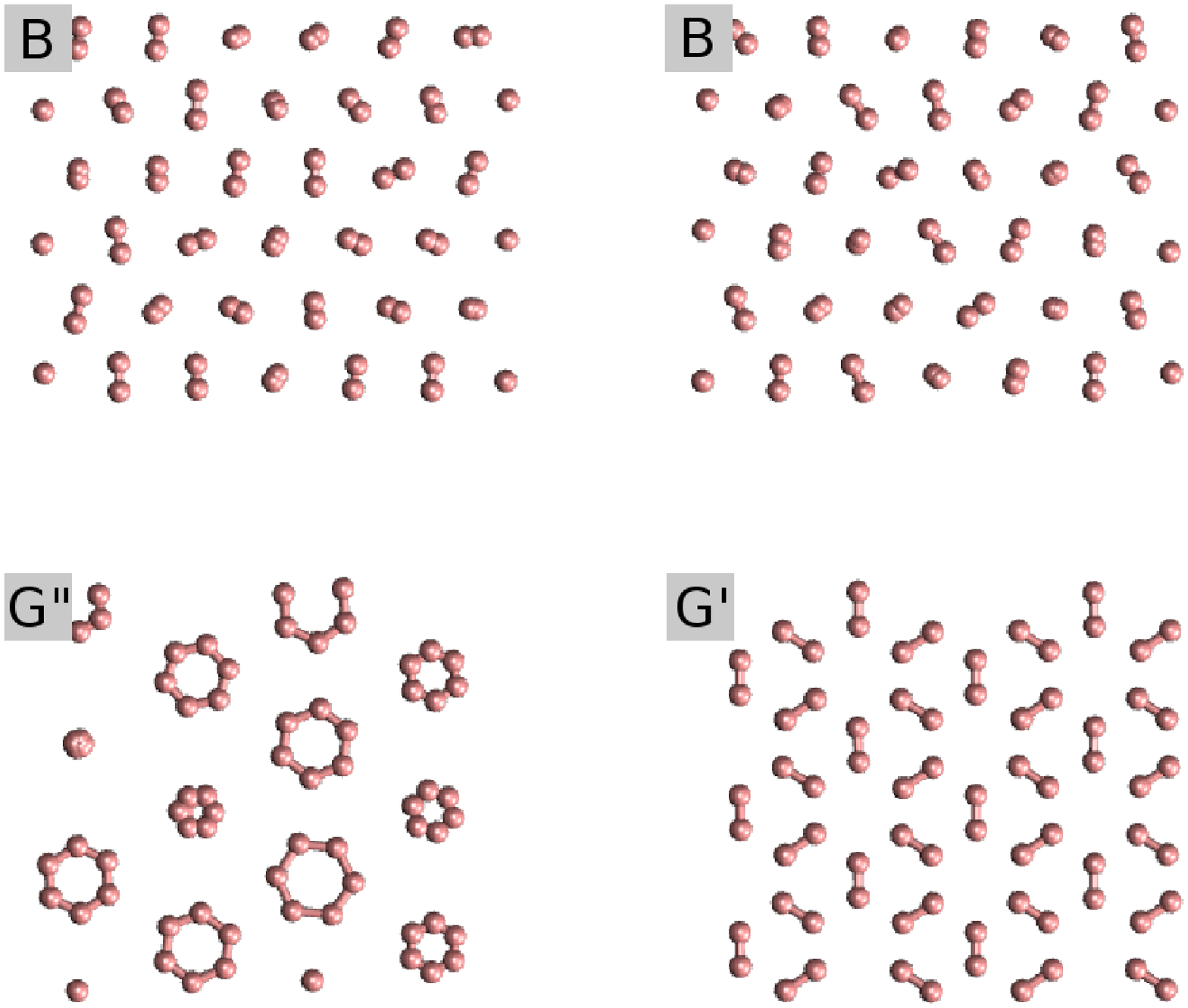}}}
\subfigure[. 300 GPa]{\fbox{\includegraphics[width=75mm]{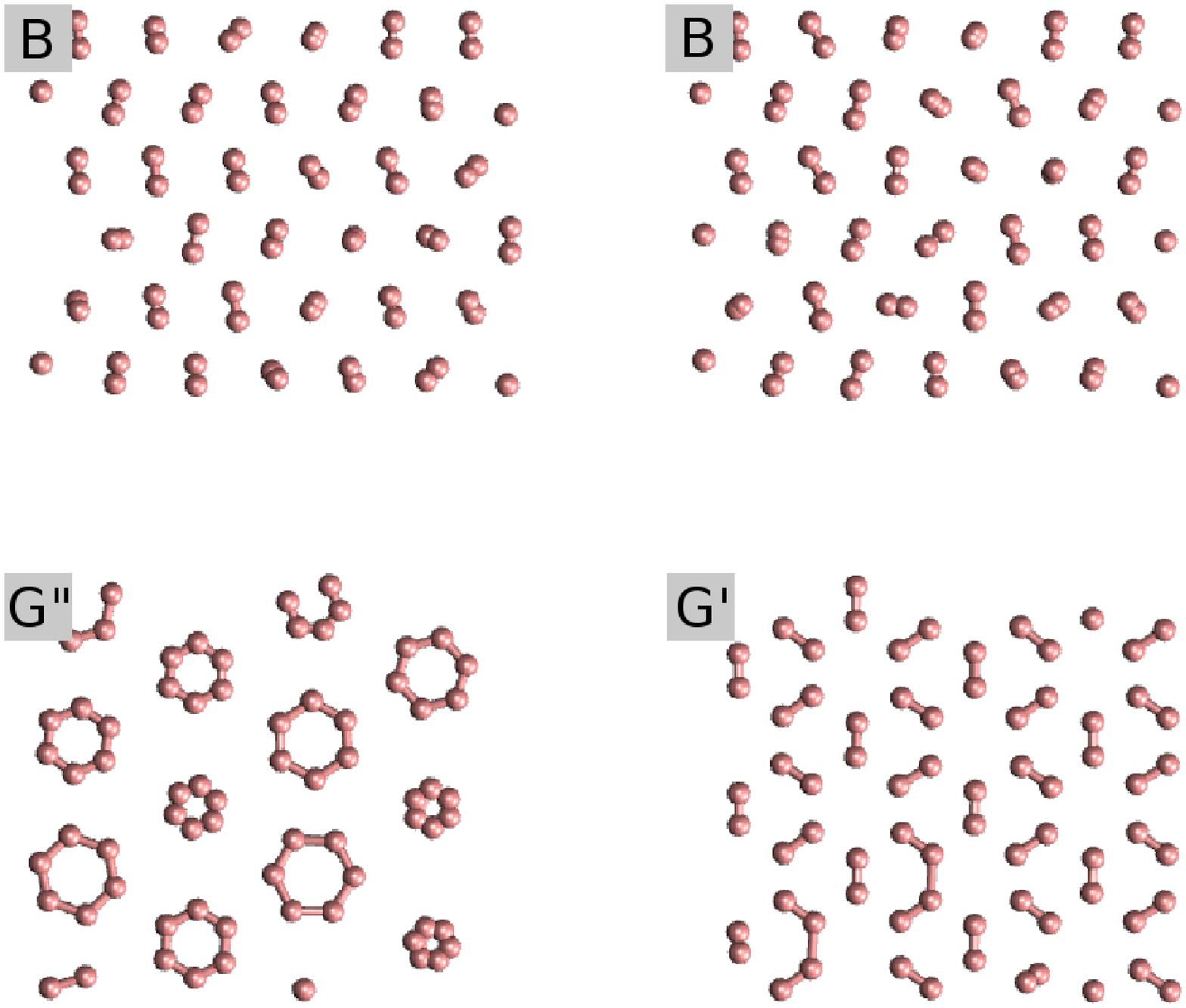}}}
\end{center}
\caption{As per figure 1 in the main paper, average positions of atoms over 1ps
at 220 K and pressures (from top left) 180 GPa, 200 GPa, 220 GPa, 250 GPa, 270 GPa, 300 GPa. Stacking is $BGBG$ and $BG'BG''$, repsectively.
\label{phaseIVX_P}}
\end{figure}

\begin{figure}[H]
\begin{center}
\includegraphics[width=150mm]{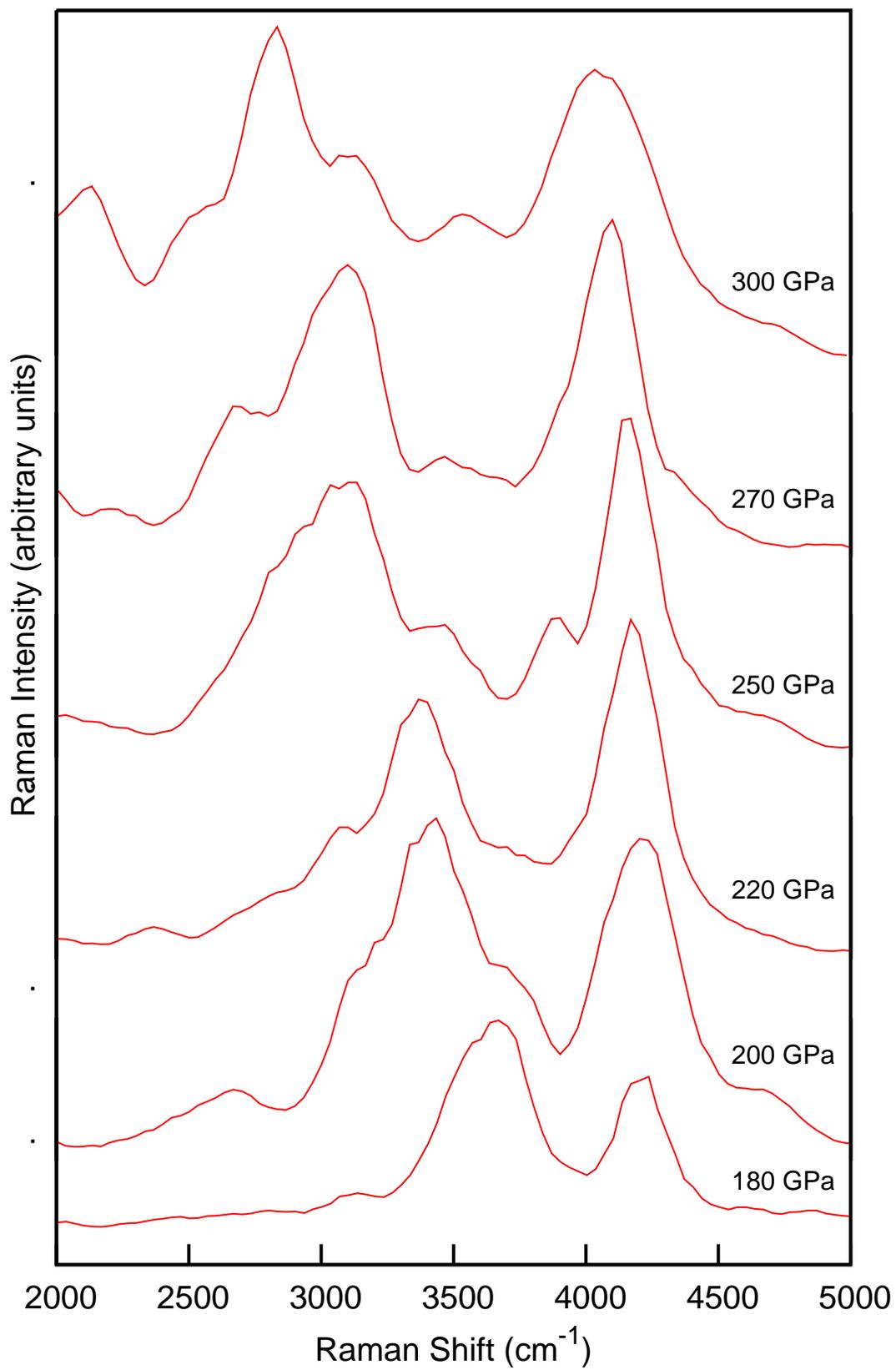}
\end{center}
\caption{Calculated Raman peaks from MD at 220 K for a range of pressures traversing the VIa-VIb transition between 250 GPa and 270 GPa.}
\label{peaks}
\end{figure}

\begin{figure}[H]
\begin{center}
\includegraphics[width=150mm]{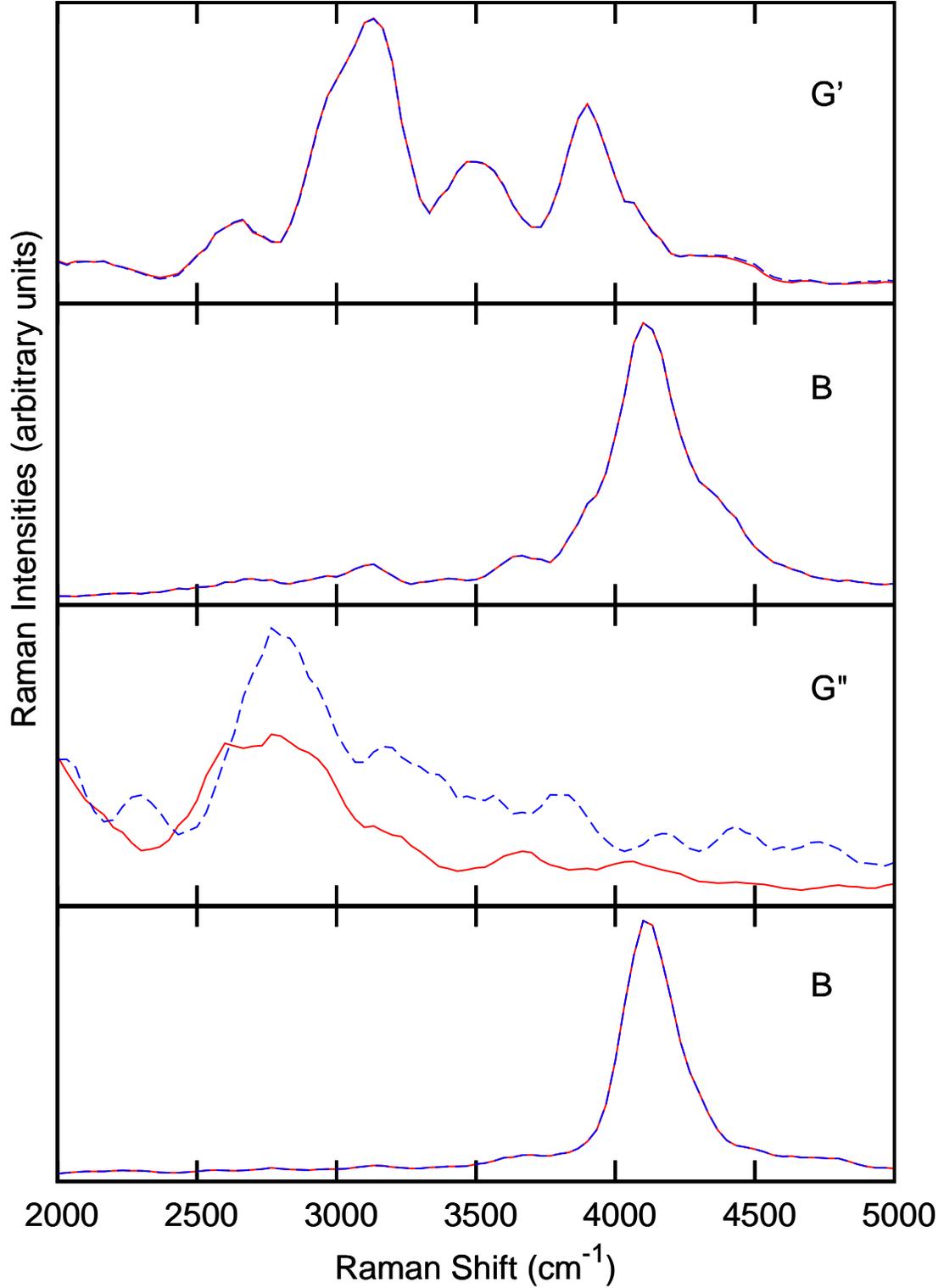}
\end{center}
\caption{Calculated Raman peaks from MD started in Pc at 220 K, 270 GPa, individually for each layer with two
different methods: identifying molecules at $t=0$ and keeping the same labels throughout the simulation (red), identifying molecules
at each time step (blue). For $B$ and $G'$ layers the same results are found, while $G''$ gives very different results depending on the
method, suggesting continous rebonding. Note that the sum of the four spectra will be slightly different from the spectrum of the whole structure,
for which the complex FT phases have to be taken into account.}
\label{PeaksLayer}
\end{figure}

\begin{figure}[H]
\begin{center}
\includegraphics[width=120mm]{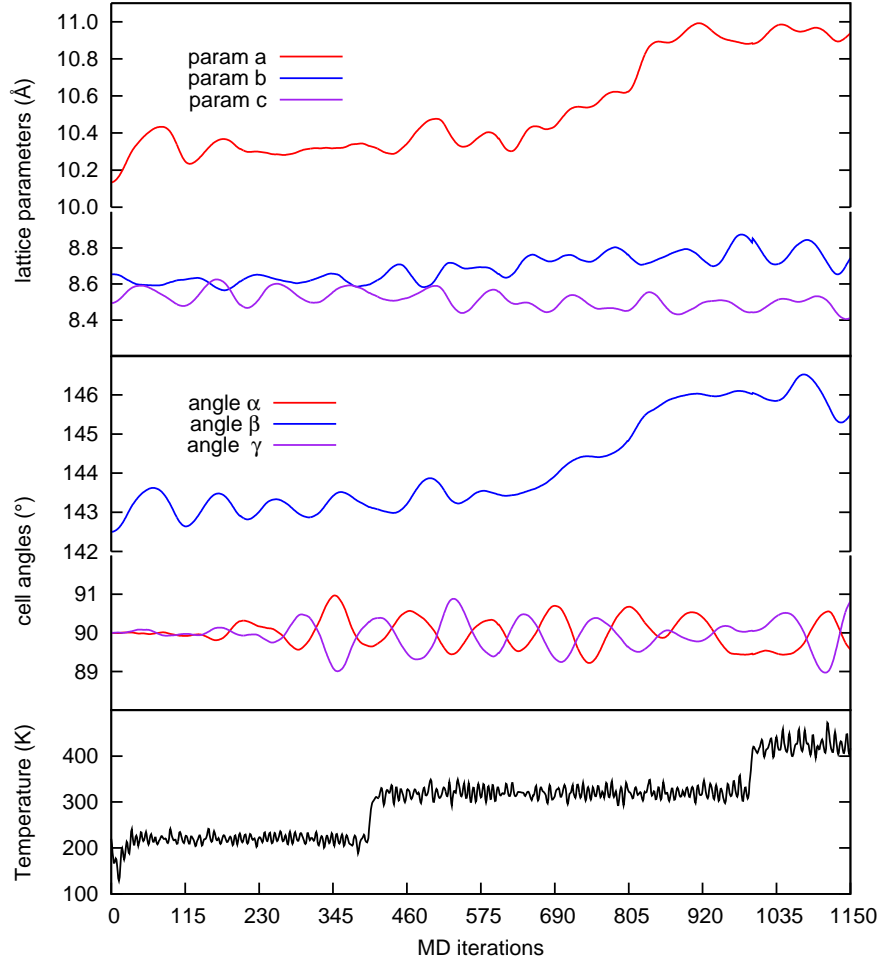}
\end{center}
\caption{Figure showing lattice parameters and angles for a long MD
  run with a ramped temperature rise, started in the $C2/c$ structure (see simulation 24 from Table \ref{MDTable} and Fig. \ref{PTo}).
  The phase transition from III-IV appears evident, however close
  comparison of Raman data with experiment reveals that the high-T
  phase is not consistent with experiment.  Parrinello-Rahman dynamics
  is not able to achieve the massive cell-shape change require to
  reach phase IVb.
\label{heatcool}}
\end{figure}

\begin{table}[H]
\centering
\begin{tabular}{|l|l|l|l|}
\hline
Phase & symmetry & conditions & description \\
\hline
gas & & ambient & molecular $H_2$\\
 I & hcp  & low-T + pressure &  quantum rotation of $H_2$ molecules.\\
 II & distorted hcp
 & Low temperature, $<$150 GPa & symmetry-breaking distortion of I\\ 
III & C2/c & 150+ GPa  300- K & layer molecules arranged in hexagonal trimers.\\
IV & & 200+ GPa, ambient temperature & hexagonal and free-rotating molecular layers \\
\hline
\end{tabular}
\caption{Summary of known properties of phases on hydrogen;
phase boundaries in hydrogen are not yet definitively established.}
\end{table}


\begin{thebibliography}{50}
\bibitem{note1} We note that convergence of the Raman signal requires a much higher $k$-point
  sampling than either structure or phonon calculations.
\bibitem{note2} It is possible to identify the layers, and to project
  only onto molecules in one layer.  Each different layer produces a
  single peak: the higher frequency peak is associated with the B
  layers.  However, there is interference between similar layers, so
  the sum of the layer projections does not give the correct Raman signal.
\end{thebibliography}
\end{document}